\documentclass[prd,a4paper,showpacs,byrevtex,preprint,11pt]{revtex4-1}
\usepackage{dcolumn}
\usepackage{amsmath}
\usepackage{array}
\usepackage{bm}
\usepackage{amssymb}
\usepackage{amsfonts}
\usepackage{epsfig}

\begin{document}

\title{The structure of the Yang-Mills spectrum for arbitrary simple gauge algebras}

\author{Fabien \surname{Buisseret}}
\thanks{F.R.S.-FNRS Postdoctoral Researcher}
\email[E-mail: ]{fabien.buisseret@umons.ac.be}

\affiliation{Service de Physique Nucl\'{e}aire et Subnucl\'{e}aire,
Universit\'{e} de Mons--UMONS,
Acad\'{e}mie universitaire Wallonie-Bruxelles,
Place du Parc 20, B-7000 Mons, Belgium}

\date{\today}

\begin{abstract}

The mass spectrum of pure Yang-Mills theory in $3+1$ dimensions is discussed for an arbitrary simple gauge algebra within a quasigluon picture. The general structure of the low-lying gluelump and two-quasigluon glueball spectrum is shown to be common to all algebras, while the lightest $C=-$ three-quasigluon glueballs only exist when the gauge algebra is A$_{r\geq 2}$, that is in particular $\mathfrak{su}(N\geq3)$. Higher-lying $C=-$ glueballs are shown to exist only for the A$_{r\geq2}$, D$_{{\rm odd}-r\geq 4}$ and E$_6$ gauge algebras. The shape of the static energy between adjoint sources is also discussed assuming the Casimir scaling hypothesis and a funnel form; it appears to be gauge-algebra dependent when at least three sources are considered. As a main result, the present framework's predictions are shown to be consistent with available lattice data in the particular case of an $\mathfrak{su}(N)$ gauge algebra within 't Hooft's large-$N$ limit.
\end{abstract}



\maketitle

\section{Introduction}

Yang-Mills (YM) theory can be in principle formulated for any gauge (or colour) algebra, although the $\mathfrak{su}(3)$ case has been maximally studied since it corresponds to the pure gauge sector of QCD. Besides the intrinsic interest of such a task, there are several motivations to study YM theory with an arbitrary gauge algebra. Let usnow give two of them. First, dealing with a generic $\mathfrak{su}(N)$ gauge algebra instead of an $\mathfrak{su}(3)$ one allows to deal with 't Hooft's large-$N$ limit of YM theory~\cite{lnco1}. This limit deserves a considerable interest nowadays due to its ability to provide relevant informations about nonperturbative phenomena in YM theory and to its central role in the celebrated AdS/CFT correspondence. Second, a change of gauge algebra allows to check deeper theoretical approaches describing $\mathfrak{su}(3)$ YM theory. For example, it has been suggested in~\cite{sve} that the phase transition of YM theory with gauge algebra $\mathfrak{g}$ might be driven by a spontaneous breaking of a global symmetry related to the center of $\mathfrak{g}$. Effective $Z_3$-symmetric models are indeed able to describe the first-order phase transition of $\mathfrak{su}(3)$ YM thermodynamics as observed in particular in lattice QCD~\cite{Z3}. However, a similar phase transition has been observed in lattice simulations of G$_2$ YM theory~\cite{G21} even though the center of G$_2$ is trivial. This shift from $\mathfrak{su}(3)$ to G$_2$ thus helps to better understand the general mechanisms of (de)confinement in YM theory by showing that the breaking of center symmetry is not the only mechanism responsible for deconfinement.

Regardless of the gauge algebra considered, a crucial feature of YM theory (even coupled with fermions) is its $\beta$-function, whose two-loop expression is well-known, see \textit{e.g.}~\cite{caswell}. In the case of pure YM, the coefficients $\beta_0$ and $\beta_1$ are strictly positive, and asymptotic freedom is present for any simple gauge algebra, on which we focus in this paper. The two-loop running coupling can then be extracted from the $\beta$-function. From its expression, one can define in the pure YM case~\cite{caswell} 
\begin{equation}\label{g2}
g^2=\frac{\lambda}{C_2^{(ad\hspace{0.1pt}j)}} \quad {\rm or}\quad \alpha_s=\frac{\alpha_0}{C_2^{(ad\hspace{0.1pt}j)}},
\end{equation}
where $\alpha_s=g^2/4\pi$ and where $\alpha_0=\lambda/4\pi$. The function $\lambda$, depending on the energy scale, has a fixed form for any gauge algebra. 

Because YM theory exhibits confinement for any nonabelian gauge algebra, it allows the existence of purely gluonic bound states called glueballs. Questions that logically arise are: What is the global structure the low-lying YM spectrum and does it strongly depend on the considered gauge algebra? The most fundamental way to study the YM spectrum is in principle by resorting to lattice computations, that have led to accurate results in the $\mathfrak{su}(3)$ case~\cite{glueb1,Chen}, but also in the case of $\mathfrak{su}(2)$~\cite{teper}, $\mathfrak{su}(8)$~\cite{Meyer}, and in the large-$N$ limit~\cite{luciN}. The YM spectrum has not been computed with other gauge algebras than $\mathfrak{su}(N)$ so far on the lattice. So it is worth addressing the problem by using an effective framework in view of getting first qualitative results and motivating future lattice computations. The quasigluon picture, in which glueballs are assumed to be bound states of a fixed number of constituent gluons, called quasigluons, will be used hereafter. Such a framework has proved to give an accurate and intuitive description of the lattice data in the $\mathfrak{su}(3)$ case, see \textit{e.g.} the review \cite{glurev}, or more specifically \cite{cons,cg3}. A quasigluon approach is actually the framework in which the gauge-algebra dependence of the observables appears the most clearly and can be the most straightforwardly dealt with. That is why it is the most appropriate framework for our purpose.  

This paper is organized as follows. The quasigluon picture is presented in Sec.~\ref{gene}. As an application of this picture, the qualitative features of the low-lying YM spectrum are discussed in Sec.~\ref{struym}: Predictions are given concerning the mass hierarchy of gluelumps and glueballs. Section~\ref{statip} is then devoted to the potential energy between static adjoint sources, which is obviously related to the instantaneous potential that could be used to build an explicit Hamiltonian in a quasigluon approach. The results obtained in Secs.~\ref{struym} and \ref{statip} are applied to the case of an $\mathfrak{su}(N)$ gauge algebra in Sec.~\ref{largen} and compared to recent lattice data. Finally, conclusions are drawn in Sec.~\ref{conclu}, while useful Lie-algebraic data are gathered in Appendix~\ref{app}. Comments on the ${\cal N}=1$ supersymmetric (SUSY) YM case are given in Appendix~\ref{susya}.

\section{The quasigluon picture}\label{gene}

\subsection{Generalities}
The YM field, or gluonic field, is defined as $A^\mu=A^\mu_a\, T_{(r)}^a$, where $\mu$ is the spacetime index, while $T_{(r)}^a$ denotes the generators of an arbitrary simple Lie algebra in the representation $r$. The charge-conjugate gluon field, $ A^{{\cal C}}_{\ \mu}$, is defined as \cite{Frit}
\begin{equation}\label{cconj}
    A^{{\cal C}}_{\ \mu}={\cal C}\, A_\mu\, {\cal C}^{-1}=-A^{{\rm T}}_\mu,
\end{equation}
where T denotes the transposition of the $T^a_{(r)}$ matrices. In $3+1$ dimensions, \textit{i.e.} the case considered in the rest of this paper, the YM theory thus contains ${\rm dim}\, ad\hspace{0.1pt}j$ transverse massless particles, or gluons -- this is also the case with more than 3 spatial dimensions. Note that $ad\hspace{0.1pt}j$ denotes the adjoint representation of the considered algebra.  

The main advantage of a quasigluon picture is that the quantum state of a given bound state made of $n_g$ quasigluons can be decomposed as $\left|n_g; \, J^{PC}\right\rangle=\left|colour \right\rangle \otimes \left|spin-space \right\rangle $, so that only the colour part of the state will qualitatively depend on the chosen gauge algebra. The spin-space part could also depend on the gauge algebra through a change in the numerical values of the parameters appearing in a specific Hamiltonian, but this would only affect quantitatively the mass spectrum, not the basic properties studied in this work like the allowed quantum numbers and the mass hierarchy. As stressed in \cite{cons}, those global properties are mostly consequences of the colour structure of a given state -- responsible for $C$ according to (\ref{cconj}) -- and of the peculiar dynamics induced by the transversality of the quasigluons and the Pauli principle -- responsible for $J^P$.

It has to be said that the idea of modelling hadrons as bound states of constituent particles is not new: The classification of baryons and mesons within the quark hypothesis is a first and successful application of such an idea. Moreover, potential approaches in which mesons and baryons are respectively seen as genuine quark-antiquark and three-quark states are able to correctly describe the mass spectra of those hadrons. We refer the interested reader to the review \cite{Lucha} for more information on this topic. A constituent picture of mesons and baryons is expected to work well for states made of heavy quarks, since in that case an expansion of QCD in inverse powers of the quark masses leads to a nonrelativistic, Schr\"odinger-like, description of those hadrons (see \textit{e.g.} the recent review \cite{brambi}). This expansion cannot be applied to hadrons made of light quarks, although potential quark models give good results in that case too. It is generally argued that the viability of quark models in this sector comes from the fact that light quarks gain a non negligible mass generated dynamically by chiral symmetry breaking \cite{szcz}: It is then justified to make a Fock-space expansion of, say, a light mesonic state and keep only its dominant, quark-antiquark, component. The situation might appear more problematic in the YM sector, where it is not so clear that a Fock-space expansion in terms of constituent gluons, or quasigluons, may be done. Although the quasigluon picture can be justified a posteriori by its ability to reproduce the $\mathfrak{su}(3)$ YM spectrum, it is worth giving some arguments justifying a priori the relevance of using such a picture to describe glueballs in generic YM theories. This is done in the next section.

\subsection{Relevance of the method}

The emergence of a constituent picture appears naturally in Coulomb gauge QCD \cite{cg0}. In this approach, the QCD Hamiltonian is written with the gluonic field in the Coulomb gauge. A key feature of this gauge is that the elimination of the nondynamical degrees of freedom creates an
instantaneous nonperturbative interaction. One can schematically write the interaction Hamiltonian as~\cite{cg0}
\begin{equation}
H_I=-\frac{1}{2}\int d{\bm x}\, d{\bm y}\, \rho^a(\bm x)\, K_I(|\bm x-\bm y|)\, \rho_a(\bm y),
\end{equation}
where $\rho^a(\bm x)$ is the colour current 
\begin{equation}
\rho^a(\bm x)=\psi^\dagger(\bm x) T^a_{(r)}\psi(\bm x)+f^a_{\ bc}\bm A^b(\bm x)\cdot\bm E^c(\bm x),
\end{equation}
and where $K_I(z)$ is the interaction kernel. 

In a first approach, the interaction kernel can be taken of funnel form, that is $K(z)=az-b/z$. Both the linear and Coulomb term are expected to be present for any gauge algebra. Indeed, the Coulomb term is the effective potential corresponding to a one-gluon-exchange diagram; one finds $b=\alpha_s$, or $b=\alpha_0/C_2^ {(ad\hspace{0.1pt}j)}$ following (\ref{g2}), independently of the considered gauge algebra. The long-range part is a linearly confining term, that can be analytically obtained by performing a strong coupling expansion of the Wilson loop in YM theory. This expansion suggests that $a=g^2 \Omega$ \cite{deldeb}, or $a\equiv\sigma_0/C_2^{(ad\hspace{0.1pt}j)}$ following (\ref{g2}), where $\Omega$ and $\sigma_0$ are some numerical parameters. Again, this result is valid for any gauge algebra, suggesting that a funnel form for the interaction kernel might be quite universal: It has at least the correct short- and long-range limits, hence it does not seem unreasonable to assume this form as a general interaction potential. Taking into account the generators of the gauge algebra present in the colour currents, the proposed forms for $a$ and $b$ lead to the static potential 
\begin{equation}\label{V2}
V_{(r)}(z)=\frac{C_2^{(r)}}{C_2^{(ad\hspace{0.1pt}j)}}\left( \sigma_0\, z-\frac{\alpha_0}{z}\right)\equiv\frac{C_2^{(r)}}{C_2^{(ad\hspace{0.1pt}j)}}\, U(z)
\end{equation}
between two sources in the representation $r$ (or $\bar r$) bound in a colour singlet. The colour factor appearing in the Coulomb term agrees with what is expected from a one-gluon-exchange process, while the total string tension is
\begin{equation}\label{sigd}
\sigma^{(r)}=\frac{C_2^{(r)}}{C_2^{(ad\hspace{0.1pt}j)}}\sigma_0.
\end{equation}
This expression agrees with the strong coupling expansion performed in \cite{deldeb} and with the Casimir scaling hypothesis, stating that a static coloured source generates a flux tube whose tension is proportional to its quadratic Casimir operator. It is well-known that the shape (\ref{V2}) is in very good agreement with lattice computations of the energy between two static coloured sources \cite{balirev}. Moreover, the Casimir scaling has been observed in many lattice simulations \cite{bali,bicu1}, including recent computations with a G$_2$ gauge algebra \cite{G2}, and is supported by effective approaches too \cite{semay}.

Using the above interaction Kernel, the gluon mass gap equation in Coulomb gauge QCD can be written as \cite{cg1}
\begin{equation}\label{mgap}
\omega(q)^2=q^2+ \int \frac{d\bm k}{4(2\pi)^3}\tilde U(|\bm k+\bm q|)\, [1+(\hat{ \bm k}\cdot\hat{ \bm q})^2]\, \frac{\omega(k)^2-\omega(q)^2}{\omega(k)},
\end{equation}
where $v=|\bm v| $, $\hat{\bm v}=\bm v/v$, and where $\tilde U$ is the Fourier transform of $U$. The resolution of (\ref{mgap}) leads to $\omega(q)^2=q^2+m_g(q)^2$, which is the dispersion relation of a quasigluon. Typically, $m_g(0)=$600-800 MeV is found in Coulomb gauge QCD for the gauge algebra $\mathfrak{su}(3)$ \cite{cg3,cg1}, and a quasigluon can then be seen as a transverse particle with a dynamically generated mass $m_g(q)$. The existence of such a mass justifies a Fock space expansion of gluonic states in terms of states with a given number of quasigluons. Eventually, only the component with the minimal number of quasigluons will be kept. As pointed out in \cite{cg0}, such a truncation makes sense because the Fock space expansion converges quickly. From dimensional analysis, the mass gap equation (\ref{mgap}) will lead to a solution of the form $m_g(q)=\sqrt{\sigma_0}\, \bar m(q/\sqrt{\sigma_0}\, ,\alpha_0)$, that does not depend qualitatively on the gauge algebra. Quantitatively however, it can be so because the parameters $\sigma_0$ and $\alpha_0$ may be gauge-algebra dependent. 

Two remarks should now be done. First, although not universally accepted, a dynamically generated gluon mass is favoured by some lattice results related to the zero momentum behaviour of the gluon propagator in Landau gauge, see \textit{e.g.} \cite{gluml1,gluml2}. Also nonperturbative field-theoretical calculations, using for example the pinch technique, find a nonzero dynamically generated gluon mass in $3+1$ YM theory \cite{glumass,glumass2}. These last results have been obtained in the case of an $\mathfrak{su}(N)$ gauge algebra. Second, it has been shown in the recent lattice study \cite{maas} that the qualitative behaviour of the Landau gauge gluon propagator does not depend on the chosen gauge algebra in the $1+1$ and $2+1$ dimensional cases. 

It is thus reasonable to conclude from the above discussion that the quasigluon picture, a priori justified by the existence of a dynamically generated gluon mass, may be used for any simple gauge algebra. 

\section{Structure of the Yang-Mills spectrum}\label{struym}
\subsection{Gluelumps}

Gluelumps are colour singlet bound states of the YM field plus a static adjoint source defined as $\phi=\phi_a\, T_{(r)}^a$. Although not ``physical" in the sense that they require the presence of an extra static source, gluelumps are nevertheless worth of interest since in QCD, $\phi$ can be seen as an adjoint static quark-antiquark pair in the limit where their separation goes to zero~\cite{gluel2}. So, the gluelump mass is grosso modo the binding energy of a heavy hybrid meson~\cite{gluel3} and, accordingly, the static source should have the quantum numbers of a static, scalar, pointlike, quark-antiquark pair: $J^{P}=0^{-}$. Hence $\eta_\phi=-1$ is assumed for the source's intrinsic parity. Moreover, $\phi^{\cal C}=\phi^T$ is taken here, implying that the charge conjugation of the lightest gluelumps will always be negative (see below). The above discussion suggests that, within a quasigluon picture, the lowest-lying gluelumps are states made of one quasigluon, the presence of the static source allowing to build a colour singlet. 

The tensor product of the adjoint representation by itself has schematically the following structure for any gauge algebra:
\begin{equation}\label{gluel}
ad\hspace{0.1pt}j\otimes ad\hspace{0.1pt}j=\bullet^S\oplus ad\hspace{0.1pt}j^A\oplus \dots ,
\end{equation}
where the $S$ ($A$) superscript denotes a(n) (anti)symmetric colour configuration, and where the singlet is represented by $\bullet$. It can moreover be checked that no algebra allows the appearance of a $\bullet^A$-term in the right hand side of (\ref{gluel}). The colour configuration of a gluelump state in the $\bullet^S$ channel is given by $\delta_{ab}\phi^a A_\mu^b\propto{\rm Tr}(\phi A_\mu)=-{\rm Tr}(\phi^{\cal C} A_{\mu}^{\cal C})$ and has always a negative charge conjugation. Now that the colour part of the gluelump wave function has been fixed thanks to~(\ref{gluel}), its spin-space wave function can be defined using Jacob and Wick's helicity formalism applied to a single transverse spin-1 particle~\cite{wick}. One finds two families of states, both with spin $J\geq 1$, among which the lightest states are obviously those with $J=1$: They are the states with the minimal rotational energy. Once decomposed in a standard $\left|^{2S+1}L_J\right\rangle$ basis, these lightest states read~\cite{gluel4}
\begin{eqnarray}
\label{hsgl1}
\left|g;1^{+-}\right\rangle &=& \left[\frac{2}{3}\right]^{1/2}\ | ^3 S_1 \rangle +
\left[\frac{1}{3}\right]^{1/2}\ | ^3 D_1 \rangle ,\\ 
\left|g;1^{--}\right\rangle & =& -| ^3 P_1 \rangle.
\end{eqnarray} 
The $1^{+-}$, being dominated by a $S$-wave component, will be the lightest, followed by the $1^{--}$.

In summary, any gauge algebra allows the existence of colour singlet states made of a single quasigluon and a static adjoint source, that are called gluelumps. These states will be at the bottom of the YM spectrum since they are made of only one quasigluon. Moreover, the spin-space part of their wave function shows that the lowest-lying gluelumps will have the quantum numbers $1^{+-}$ and $1^{--}$, their masses being ordered as $M_{1^{+-}}<M_{1^{--}}$. A typical gluelump mass scale can be defined as
\begin{equation}\label{ml}
M_l=\frac{M_{1^{+-}}+M_{1^{--}}}{2},
\end{equation} 
and will be useful in the following. 

Interpreting gluelumps as one-quasigluon states agrees with $\mathfrak{su}(3)$ lattice calculations, showing that the lowest-lying gluelumps are lighter than all the currently known glueballs~\cite{gluel2,gluel1}: A minimal number of two quasigluons is indeed needed to build a glueball state (see next section). Furthermore, the mass hierarchy $M_{1^{+-}}<M_{1^{--}}$ has been found on the lattice~\cite{gluel2,gluel1}, and explicit Coulomb gauge calculations confirm that the $S$-wave dominated state is indeed the lightest one \cite{gluel4}. Note that $M_l\approx 1$~GeV according to the results of~\cite{gluel2}.

\subsection{Two-quasigluon glueballs}

In the framework used here, glueballs are colour singlet states made of quasigluons only. As shown by (\ref{gluel}), the minimal required number of constituent gluons is two, \textit{i.e.} a state whose colour configuration is $\delta_{ab}A_\mu^a A_\nu^b\propto{\rm Tr}(A_\mu A_\nu)$ $={\rm Tr}(A_{\mu}^{\cal C} A_{\nu}^{\cal C})$. Consequently, the charge conjugation of a two-quasigluon glueball is always positive. Since the colour wave function is symmetric, the Pauli principle demands the symmetrization of the spin-space wave function. Again, the building of such a state can be achieved by resorting to Jacob and Wick's helicity formalism in the case of a symmetrized state made of two transverse spin-1 particles. As computed in \cite{glueb0}, one finds four families of states that will not explicitly be written here for the sake of simplicity. It is nevertheless worth saying that a look at their decomposition in a $\left|^{2S+1}L_J\right\rangle$ basis immediately suggests that the lightest states are the scalar, tensor, and pseudoscalar ones, reading \cite{glueb0}
\begin{eqnarray}\label{scas}
    \left|gg;0^{++}\right\rangle&=&\left[\frac{2}{3}\right]^{1/2}\left|^1 S_0\right\rangle+\left[\frac{1}{3}\right]^{1/2}\left|^5 D_0\right\rangle,\\
   \left|gg;2^{++}\right\rangle&=&\left[\frac{2 }{5}\right]^{1/2} \left|^5S_2\right\rangle+\left[\frac{4}{7}\right]^{1/2}\left|^5D_2\right\rangle \nonumber\\&&+\left[\frac{1}{35}\right]^{1/2}\! \! \left|^5 G_2\right\rangle, \\
    \left|gg;0^{-+}\right\rangle&=&-\left|^3P_0\right\rangle.
\end{eqnarray} 
The $0^{++}$ state (\ref{scas}), being dominated by its $S$-wave component, is expected to be the lightest one, followed by the $2^{++}$ and $0^{-+}$ states. Explicit values of these glueball masses can be computed within a quasigluon picture using the interaction potential (\ref{V2}); it has been done in Coulomb gauge QCD by using parameters fitted on the $\mathfrak{su}(3)$ lattice data \cite{cg3,cg1} and the results agree with the mass hierarchy expected from the orbital angular momentum content of each glueball state.  

It is readily concluded that any gauge algebra allows the existence of two-quasigluon glueballs in a symmetric colour singlet, that will stand at the bottom of the glueball spectrum and should have a typical mass of $2\, M_l$. This estimate is an immediate consequence of the quasigluon picture developed in this work. Moreover, the lightest two-gluon glueballs will have the quantum numbers $0^{++}$, $2^{++}$ and $0^{-+}$, the masses being such that $M_{0^{++}}<M_{2^{++}},\, M_{0^{-+}}$. Notice an important feature of two-quasigluon states: Yang's theorem forbids the existence of two transverse spin-1 particles in a $J=1$ state~\cite{Yang}, \textit{i.e.} no vector glueball is expected around $2\, M_l$, regardless of the considered gauge algebra. 

As an illustration, it can be remarked that the lightest gluelump masses are $M_{1^{+-}}=0.87(15)$~GeV and $M_{1^{--}}=1.25(16)$~GeV~\cite{gluel2}, while the lightest glueball masses are $M_{0^{++}}=1.730(50)(80)$~GeV, $M_{2^{++}}=2.400(25)(120)$~GeV and $M_{0^{-+}}=2.590(40)(130)$~GeV in $\mathfrak{su}(3)$ lattice QCD \cite{glueb1} -- no $1^{P+}$ state is observed in this mass range. The estimate $2\, M_l$ and the hierarchy $M_{0^{++}}<M_{2^{++}},\, M_{0^{-+}}$ which are found for two-quasigluon states compare thus well with well-known lattice data. 

\subsection{Three-quasigluon glueballs}

It is worth going a step further and discuss the properties of glueballs made of three quasigluons, that is three quasigluons in a colour singlet. This is the simplest case in which the results will depend on the considered gauge algebra, as shown by the tensor product
\begin{equation}\label{3ga}
ad\hspace{0.1pt}j\otimes ad\hspace{0.1pt}j\otimes ad\hspace{0.1pt}j=\left\lbrace 
\begin{matrix}
\bullet^A &  \oplus & \dots & \\
\bullet^A & \oplus & \bullet^S &\oplus\dots\quad {\rm only\ for\ A}_{r\geq 2} 
\end{matrix}
\right. .
\end{equation}
This means that a totally antisymmetric colour singlet can always be formed by using the structure constants; one gets the colour structure $f^{abc}A_a^\mu A_b^\nu A_c^\rho\propto{\rm Tr}\left( \left[A^\mu,A^\nu\right]A^\rho\right)$ $={\rm Tr}\left( \left[A^{{\cal C}\, \mu},A^{{\cal C}\, \nu}\right]A^{{\cal C}\, \rho}\right) $ that has $C=+$. But, a peculiar feature of the algebras A$_{r\geq 2}$ is that they possess a totally symmetric invariant tensor whose indices run in the adjoint representation, generally denoted $d^{abc}$, that allows to build the totally symmetric colour singlet $d^{abc}A_a^\mu A_b^\nu A_c^\rho\propto{\rm Tr}\left( \left\{A^\mu,A^\nu\right\}A^\rho\right) =-{\rm Tr}\left( \left\{A^{{\cal C}\, \mu},A^{{\cal C}\, \nu}\right\}A^{{\cal C}\, \rho}\right) $. Such states always have $C=-$, and their phenomenological relevance is considerable since A$_r$ in its compact real form is nothing else than the algebra $\mathfrak{su}(r+1)$. 

Pauli's principle states that to a totally (anti)symmetric colour singlet must correspond a totally (anti)symmetric spin-space wave function. Although the explicit form of (anti)symmetrized three-quasigluon states within Wick's three-body helicity formalism \cite{w3b} will not be given here for the sake of clarity, some observations can be made concerning the quantum numbers of the lightest three-quasigluon states. Group-theoretical arguments developed in~\cite{cons} show that the lowest-lying three-quasigluon glueballs will have the quantum numbers $0^{-+}$ and $(1,3)^{+-}$. The absence of low-lying $0^{P-}$ state is in agreement with~\cite{fumi}, that is actually an extension of Yang's theorem to three-photon states, showing that no (pseudo)scalar three-photon state may exist. 

In summary, the lowest-lying three-quasigluon states should have a mass around $3\, M_l$. At such a mass scale one expects both excited two-gluon and low-lying three-gluon states to coexist (and probably significantly mix) in the $C=+$ sector. There, only $1^{P+}$ states could safely be interpreted as three-gluon ones. No glueball state around $3\, M_l$ is present in the $C=-$ sector excepted when the gauge algebra is A$_{r\geq2}$. 

An important check of the above discussion is that the $1^{+-}$ and $3^{+-}$ states have indeed been observed in $\mathfrak{su}(3)$ lattice QCD with a mass of $2.940(30)(140)$~GeV and $3.550(40)(170)$~GeV respectively \cite{glueb1}, while no such states exist when $\mathfrak{su}(2)$ (the compact real form of ${\rm A}_1$) is used~\cite{teper}. 

\subsection{Higher-lying states}
Obviously, higher-lying states with \textit{e.g.} four or more quasigluons may exist, but their exhaustive study would be rather tedious and will not be pursued in the present work. An important remark has nevertheless to be done: If all the representations of a given gauge algebra are real, one has $T_a^{(r)}=-(T_a^{(r)})^{\rm T}$ and $A_\mu=A_\mu^{{\cal C}}$. Then, an arbitrary colour structure $\Theta^{a_1\dots a_n}A^{\mu_1}_{a_1}\dots A^{\mu_n}_{a_n} $ has always a positive charge conjugation. Following~\cite[Chapter 15]{corn}, the algebras A$_1$, B$_{r\geq2}$, C$_r$, D$_{{\rm even}-r\geq4}$, E$_7$, E$_8$, F$_4$, and G$_2$ have only real representations; consequently, no $C=-$ glueballs can be built in those cases. 

A $n_g$-quasigluon state will have a mass around $n_g\, M_l$ in the present formalism. Such high-lying glueballs would be particularly difficult to study on the lattice because they would fall in a mass range where many-glueball states are present. It is nevertheless possible that the $0^{+-}$ glueball, observed in $\mathfrak{su}(3)$ lattice computations~\cite{glueb1,Chen}, might be interpreted as a four-quasigluon state as suggested in~\cite{cons}. 

In the particular case of large-$N$ QCD, it is known that baryons are $N$-quark states~\cite{lnco2}, that is the minimal number of quarks needed to build a totally antisymmetric colour-singlet. Studying the properties of such states demands a particular care since their number of constituents tend towards infinity in 't Hooft's limit. A similar situation does not occur for glueballs because, for any gauge-algebra, symmetric and antisymmetric colour singlets can be built with only two or three quasigluons. So, introducing states with a number of quasigluons tending toward infinity does not seem relevant from a physical point of view, and such states will not be discussed here. 
 
\section{The static potential}\label{statip}
\subsection{Two sources}
The static energy between coloured sources is an observable that is both accurately computable on the lattice and relevant in view of understanding the structure of confinement, see \textit{e.g.} the review \cite{balirev}. The simplest situation is the potential energy of two static sources in the representation $r$ (or $\bar r$) bound in a colour singlet, for which the potential energy is given by (\ref{V2}) where $z$ is the separation between the sources. When $r=ad\hspace{0.1pt}j$, one has
\begin{equation}
V_{(ad\hspace{0.1pt}j)}(z)=\sigma_0 z-\frac{\alpha_0}{z},
\end{equation} 
so the values of $\sigma_0$ and $\alpha_0$, that cannot be guessed from the present approach, could be measured thanks to lattice computation of the static energy between two adjoint sources for different gauge algebras.

\subsection{Three sources}
A particularly interesting case is that of the static energy between three adjoint sources. It can be read from (\ref{gluel}) and (\ref{3ga}) that the following colour structure exists for any algebra: $\left[ [ad\hspace{0.1pt}j,ad\hspace{0.1pt}j]^{ad\hspace{0.1pt}j^A},ad\hspace{0.1pt}j\right]^{\bullet^A} $, \textit{i.e.} each pair is in the adjoint representation, while the three sources are in an antisymmetric colour singlet. Assuming that each source generates an adjoint flux tube, the static energy corresponding to the above colour configuration should be given by the potential 
\begin{equation}\label{vy}
V_Y=\sum^3_{i=1}\sigma_0\, |{\bm r}_i-{\bm Y}|- \frac{1}{2}\sum^ 3_{i<j=1}\frac{\alpha_0}{|{\bm r}_i-{\bm r}_j|},
\end{equation}
where the source's positions are denoted ${\bm r}_i$ and whose confining part is often called Y-junction in the literature. Again, the interaction has been splitted in a short-range Coulomb part and in a long-range linear part. The point ${\bm Y}$ where the flux tubes meet is such that the sum $\sum^3_{i=1} |{\bm r}_i-{\bm Y}| $ is minimal. A confinement of Y-junction type is supported by lattice computations of the static energy between three sources in the fundamental representation of $\mathfrak{su}(3)$, especially when those sources are in a spatially symmetric configuration like an equilateral triangle~\cite{yshape}. 

The Y-junction potential is not the only allowed possibility. Excepted for E$_8$ indeed, the adjoint representation is not the lowest-dimensional one, that is called fundamental in the present work. It can be checked that the adjoint representation appears in the tensor product $f\otimes f$ when the algebra is self-dual, and in the tensor product $f\otimes\bar f$ when the algebra is not self-dual, that is for A$_{r\geq 2}$ and E$_6$. This means that an adjoint source can always generate two fundamental (or a fundamental and a conjugate) flux tubes instead of an adjoint one, E$_8$ excepted. In the case where fundamental flux tubes are present, taking into account that $C_2^{(f)}= C_2^{(\bar f)}$, the potential will be referred to as a $\Delta$-potential, whose form is
\begin{equation}\label{vd}
V_\Delta=\sum^3_{i<j=1}\left[\frac{C_2^{(f)}}{C_2^{(ad\hspace{0.1pt}j)}} \sigma_0 |{\bm r}_i-{\bm r}_j| - \frac{1}{2}\frac{\alpha_0}{|{\bm r}_i-{\bm r}_j|}\right].
\end{equation}

\begin{figure}[ht]
\begin{center}
\includegraphics[width=7cm]{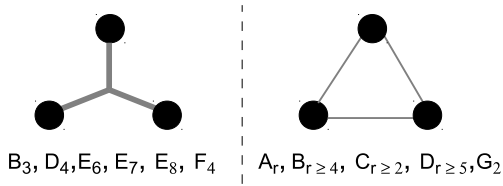}
\end{center}
\caption{Schematic illustration of the two possible confining terms for three adjoint sources located on the apices of an equilateral triangle. Left panel: Three adjoint flux tubes forming a Y-junction. Right panel: Three fundamental flux tubes forming a $\Delta$-configuration. Below are indicated the algebras for which each configuration is energetically preferred assuming the Casimir scaling hypothesis. }
\label{static}
\end{figure}

The question of knowing whether $V_Y$ or $V_\Delta$ is energetically preferred for a given gauge algebra is relevant in view of getting a better understanding of confinement in YM theory. A simple configuration is when the three adjoint sources are located on the apices of an equilateral triangle. One can then compute that, when only the confining part of the static energy is kept,
\begin{equation}\label{rat}
\frac{V_Y}{V_\Delta}=\frac{1}{\sqrt 3}\frac{C_2^{(ad\hspace{0.1pt}j)}}{C_2^{(f)}}.
\end{equation}
If this ratio is $>1$ ($<1$), $V_\Delta$ ($V_Y$) is energetically favoured. Using the data gathered in Table \ref{tab}, one can thus decide which confining term is favoured. The results are illustrated in Fig. \ref{static}. According to the value of the ratio (\ref{rat}), the gauge algebras for which the Y-junction is maximally favoured are E$_7$ and E$_8$, while a $\Delta$-shape is maximally favoured in the case of A$_1$, A$_2$ and C$_2$.

The static energy between three adjoint sources has been computed in $\mathfrak{su}(3)$ lattice QCD in~\cite{bicu2}. It appears that the long-range part of the static energy is rather compatible with $V_\Delta$ in the considered cases, namely when the sources form an equilateral or an isosceles triangle. Here we find that $V_Y/V_\Delta>1$ for the A$_r$-family, and in particular for the gauge algebra $\mathfrak{su}(3)$ in agreement with the results of \cite{bicu2,gluphen}. Moreover, as lattice results for the static energy between two coloured charges are already available for the gauge algebra G$_2$ \cite{G2}, it is reasonable to think that the picture of three-body confinement developed here will be testable in a near future.

\section{The case of $\mathfrak{su}(N)$}\label{largen}

An important result of the present paper is the prediction of the behaviour of glueball masses versus $N$ in the $\mathfrak{su}(N)$ (A$_{N-1}$) case, and especially its limit at large $N$ and constant 't Hooft coupling $\lambda=g^2N$~\cite{lnco1,lnco2}. From a phenomenological point of view indeed, calculations at leading order in $1/N$ can shed valuable insights on nonperturbative phenomena in gauge theories, even for values of $N$ as small as 3. The large-$N$ limit should in particular provide an accurate description of glueball spectroscopy in QCD since glueball masses are of order 1 in that limit \cite{lnco2}, while the first corrections should arise at order $1/N^2$ . 

\subsection{Two-quasigluon glueballs}
There have been many attempts to compute glueball masses on the lattice and see whether the mass of a given state behaves as 
\begin{equation}\label{mfit}
M_G(N)=M_G(\infty)+\frac{c_G}{N^2}
\end{equation}
or not; the interested reader will find many references about that topic in \cite{teper,luciN,luci4}. In the recent work \cite{luciN} in particular, it has been checked that the glueball spectrum obtained on the lattice is accurately described by the form (\ref{mfit}) and that the value of $c_G$ is compatible with zero (up to the error bars) for the lowest-lying glueballs. 

Is a zero value of $c_G$ compatible with the present approach? As discussed in Sec. \ref{gene}, a powerful approach relying on this picture is Coulomb gauge QCD. In that approach, the dynamically generated gluon mass, given by (\ref{mgap}), does not depend on $N$ since, by definition of 't Hooft's limit, $\sigma_0$ and $\alpha_0$ are constant with respect to $N$. Indeed, $\alpha_0=\lambda/4\pi$ and $\sigma_0=\lambda\Omega$ from (\ref{sigd}), $\Omega$ and $\lambda$ being independent of $N$. The same argument applies to the interaction potential (\ref{V2}) between two quasigluons. Consequently, nothing in the two-body part of an explicit Hamiltonian will depend on $N$, in agreement with a value of $c_G$ compatible with zero. As an illustration, one can check the remarkable independence on $N$ of the masses of the lightest scalar, pseudoscalar and tensor glueballs, corresponding to the the $A_1^{++}$, $A_1^{-+}$, and $(E,T_2)^{++}$ channels respectively, in Figs. 10, 11, 13 and 17 of \cite{luciN}. 

\begin{figure}[t]
\begin{center}
\includegraphics[width=8.0cm]{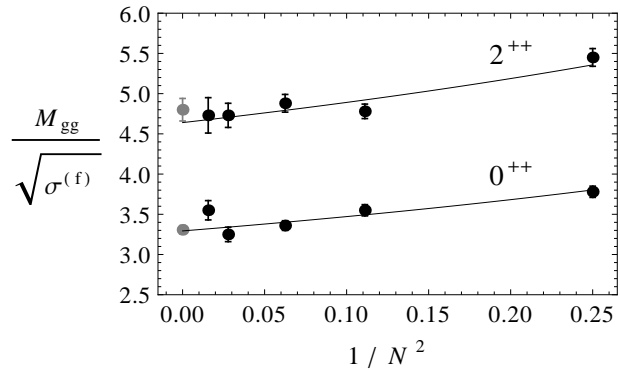}
\end{center}
\caption{Lightest scalar and tensor glueball masses normalized to the fundamental string tension computed on the lattice (black points) for different number of colours $N$ \cite{luci4}. The extrapolation to infinite $N$ has been also indicated (gray points) \cite{luci4}. The lattice data are compared to formula (\ref{mgg}) with $\theta_{gg}=2.33$ for the scalar glueball and $\theta_{gg}=3.28$ for the tensor glueball.}
\label{2g}
\end{figure}

Another check of the quasigluon picture in the two-quasigluon sector is provided by the earlier data of \cite{luci4}. In this last work, among other results, the scalar and tensor glueball masses are computed for different values of $N$ but normalized to $\sqrt{\sigma^{(f)}}$. Since the $0^{++}$ and $2^{++}$ masses should be independent of $N$ in a constituent picture, all the $N$-dependence will be contained in the normalization factor. Using (\ref{sigd}), one expects, for a two-quasigluon state, 
\begin{equation}\label{mgg}
\frac{M_{gg}}{\sqrt{\sigma^{(f)}}}=\sqrt{\frac{2N^2}{N^2-1}}\ \theta_{gg}.
\end{equation}
As shown in Fig. \ref{2g}, the above formula compares favourably to the lattice data of \cite{luci4}; the value $\theta_{gg}=2.33$ used in the plot to fit the scalar glueball masses is not unphysical since it leads to a quite low but acceptable scalar glueball mass of 1.56~GeV with $N=3$ and the standard value $\sigma^{(f)}=0.2$~GeV$^2$. Similarly, $\theta_{gg}=3.28$ leads to a standard mass of 2.20~GeV for the tensor glueball.

\subsection{Three-quasigluon glueballs}
\begin{figure}[t]
\begin{center}
\includegraphics[width=8.5cm]{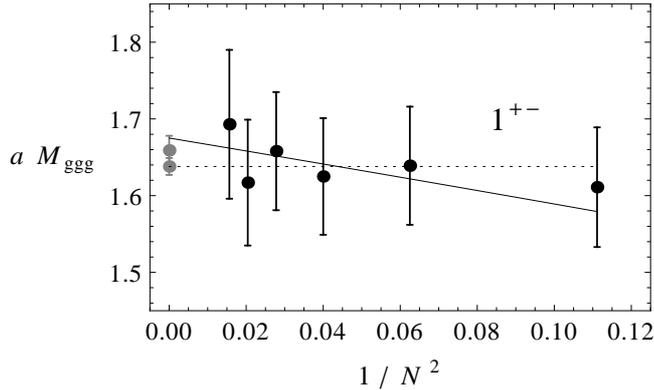}
\end{center}
\caption{Lightest vector glueball mass in units of the lattice spacing computed on the lattice (black points) for different number of colours $N$ \cite{luciN}. The two inequivalent extrapolations to infinite $N$ have been also indicated (gray points) \cite{luciN}. The lattice data are compared either to a constant mass $a\,  M_{ggg}=1.64$ (dotted line) or to formula (\ref{mggg}) with $a\, \theta_{ggg}=2.37$.}
\label{3g}
\end{figure}

In the case of a three-quasigluon glueball, the dynamically generated gluon mass remains $N$-independent, but following the previous section, the static potential between three quasigluons can be either $N$-independent, $V_Y$, or $N$-dependent, $V_\Delta$, since $C_2^{(f)}/C_2^{(ad\hspace{0.1pt}j)}=(N^2-1)/(2N^2)$. The dependence (or not) on $N$ arises at the level of the confining term, which contains the only dimensioned parameter of the system, that is the string tension. So the mass of a three-quasigluon state should be either constant if $V_Y$ is used, or of the form
\begin{equation}\label{mggg}
M_{ggg}^\Delta=\sqrt{\frac{N^2-1}{2N^2}}\, \theta_{ggg}
\end{equation}
if $V_\Delta$ is used. This last case is a priori favoured for the gauge algebra $\mathfrak{su}(N)$, as discussed in the previous section. Interestingly, the evolution of the $1^{+-}$ glueball mass, which has to be seen as the lightest three-quasigluon state, versus $N$ has been computed in \cite{luciN} (it corresponds to the $T_1^{+-}$ channel). The corresponding lattice data are plotted in Fig. \ref{3g} for $N>2$ since this state is absent for $N=2$. It appears that both a constant mass and the mass predicted by (\ref{mggg}) are compatible with the current error bars. In the case where the fitting form (\ref{mfit}) with $c_G\neq 0$ is assumed, it is found from the aforementioned lattice data that $a\, M_{1^{+-}}=1.659(19)-0.4(0.3)/N^2$ \cite{luciN}, while an expansion of formula (\ref{mggg}) leads to $a\, M_{1^{+-}}=1.68-0.84/N^2$, that is a quite similar behaviour. This in an indication that further lattice calculations in the $C=-$ glueball sector could be very useful in order to disentangle the different models of confinement.

\section{Conclusions}\label{conclu}

This work aimed at studying the structure of the low-lying mass spectrum of pure $3+1$-dimensional Yang-Mills theory with a generic simple Lie algebra as gauge algebra. Purely gluonic bound states are indeed expected since Yang-Mills theory exhibits asymptotic freedom (and thus confinement) as soon as the gauge algebra is nonabelian. 

\begin{figure}[t]
\begin{center}
\includegraphics[width=6cm]{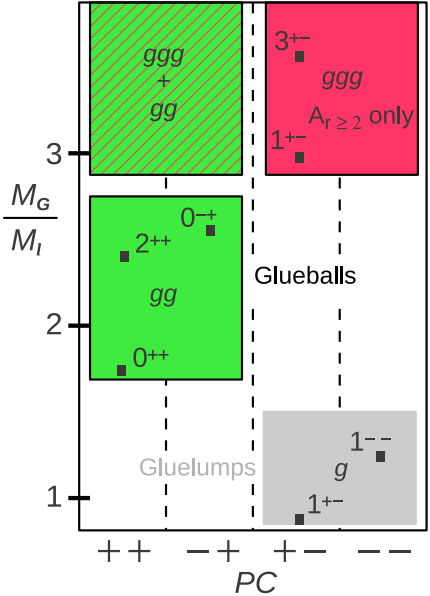}
\end{center}
\caption{(Color online) Schematic representation of the Yang-Mills spectrum for arbitrary simple gauge algebras. $M_G$ denotes the mass of a given state, while $M_l$ is the typical mass of the lightest gluelumps. Results from $\mathfrak{su}(3)$ lattice QCD have been indicated for comparison (squares) \cite{glueb1,gluel2}; the error bars are not shown for the sake of clarity.}
\label{spec}
\end{figure}

Informations about glueball and gluelump masses and quantum numbers have been obtained within a quasigluon picture: Gluonic bound states have been modelled as systems made of a given number of transverse adjoint particles called quasigluons. Such a framework, that has already proven to be relevant in QCD, might be justified thanks to the phenomenon of dynamical gluon mass generation, expected to occur for any gauge algebra. The dynamically generated mass justifies a Fock space expansion of a given gluonic state in terms of quasigluons interacting via a static potential \cite{cg0,cg1}. From this starting point it has been shown that the lightest possible states are the $1^{+-}$ and $1^{--}$ gluelumps, seen as one-quasigluon states plus a static colour adjoint source. At about two times the typical gluelump mass, $M_l$, appear two-quasigluon glueballs in the $C=+$ sector, the lightest of which is the $0^{++}$, followed by the $2^{++}$ and $0^{-+}$. At masses similar to $3\, M_l$, three-quasigluon bound states should be present. In general, they can only have $C=+$ and thus probably mix with excited two-gluon states. In the special case of an A$_{(N-1)\geq 2}$ gauge algebra, \textit{i.e.} $\mathfrak{su}(N\geq3)$ in particular, $C=-$ three-quasigluon states may also be built and are indeed observed on the lattice. A summary plot of those results is shown in Fig.~\ref{spec}, where it can be checked that our results compare well with lattice data in the $\mathfrak{su}(3)$ case. 

Not only the gluonic spectrum may be affected by the chosen gauge algebra, but also the static energy between coloured sources. Assuming the Casimir scaling hypothesis, a long-range linear potential with adjoint string tension is always expected between two adjoint sources. However, in the case of three adjoint sources, either a confinement of Y-shape with three adjoint strings or a $\Delta$-shape with three fundamental strings is allowed. In the case of sources forming an equilateral triangle, is has been shown that the $\Delta$-shape is energetically favoured, excepted for the algebras B$_3$, D$_4$, E$_{6,7,8}$, and F$_4$. Such a prediction could be testable in a near future using lattice calculations.

The main result of this work is that, by particularizing the present framework to the case of an $\mathfrak{su}(N)$ gauge algebra within 't Hooft's limit, the behaviour of the low-lying glueball masses versus the number of colours can be predicted and favourably compared to recent lattice data. The observed independence of the masses with respect to $N$ is naturally explained for two-quasigluon glueballs. For three-quasigluon states, the mass can be either constant or slightly increasing with $N$ following that the Y- or $\Delta$-shape is used as static potential. Although the $\Delta$-shape is energetically favoured, the current accuracy of the lattice data does not allow to exclude one case or another. 

The allowed quantum numbers for gluelumps and glueballs could be recovered in a model-independent way by studying \textit{e.g.} the various gluelump- and glueball-generating field-strength correlators for any gauge algebra, similarly to what is done in~\cite{jaffe}. Discussing the mass hierarchy of YM bound states demands however to take into account some dynamical information, which is here provided by the quasigluon approach. So the present results are quite general since they are formulated for an arbitrary gauge algebra, but have to be seen as model-dependent in the sense that a particular way of modelling quasigluon dynamics has been used.

As an outlook, it is worth mentioning that the structure of the complete glueball spectrum (not only the lowest-lying states) with different gauge algebras might help to clarify the phase diagram of Yang-Mills theory. For example, the nature of the phase transition in finite-temperature $\mathfrak{su}(N)$, $\mathfrak{sp}(2)$, and E$_7$ Yang-Mills theory has been studied in the recent work \cite{braun} by using renormalization-group methods and a Polyakov-loop-based approach. A first-order phase transition has been found in all cases after computation of the Polyakov loop. One might then compute the equation of state of Yang-Mills theory in the confined phase by identifying it to the one of a glueball gas, as it has been successfully done in \cite{hage} for the gauge algebra $\mathfrak{su}(3)$. Another outlook is the addition of one flavour of massless adjoint fermions, leading to a ${\cal N}=1$ supersymmetric Yang-Mills theory~\cite{salam}, which still enjoys asymptotic freedom for any gauge algebra. Some comments about this case are given in Appendix~\ref{susya}, but a detailed study is left for future works
      
\section*{Acknowledgments}    
I thank the F.R.S.-FNRS for financial support, N. Boulanger for fruitful discussions about the present manuscript, and C. Semay for his careful reading of the paper.  

\begin{appendix}
\section{Lie-algebraic data}\label{app}

Here are gathered some data concerning simple Lie algebras, following the conventions of~\cite{fuchs}. Let $\mathfrak{g}$ be a simple Lie algebra defined by the commutation relations
\begin{equation}
[T^a,T^b]=f^{ab}_{\phantom{ab}c}\, T^c,
\end{equation}
where $T^a$ denote the generators of $\mathfrak{g}$. The generators of $\mathfrak{g}$ in the adjoint representation are then given, in matrix form, by $(T_{(ad\hspace{0.1pt}j)}^a)^b_c=f^{ab}_{\phantom{ab}c}$. The Cartan-Killing form, $\kappa^{ab}$, and its inverse, $\kappa_{ab}$, are respectively defined by 
\begin{equation}
\kappa^{ab}=\frac{1}{I_2^{(ad\hspace{0.1pt}j)}}{\rm Tr}\left( T_{(ad\hspace{0.1pt}j)}^a T_{(ad\hspace{0.1pt}j)}^b\right)  \quad{\rm and}\quad \kappa^{ab}\kappa_{bc}=\delta^ a_c,
\end{equation}
where $I_2^{(ad\hspace{0.1pt}j)}$ is the second-order Dynkin index of the adjoint representation. Of particular interest for the present work is the quadratic Casimir operator
\begin{equation}\label{c2d}
\hat C_2=\kappa_{ab}\, T^a\, T^b.
\end{equation}
Its eigenvalue in the adjoint representation reads (the highest root is normalized to 1)
\begin{equation}
C_2^{(ad\hspace{0.1pt}j)}=g^V,
\end{equation}
where $g^V$ is the dual Coxeter number of the considered algebra, while its eigenvalue in any representation can be computed by using the formula
\begin{equation}
C_2^{(r)}=\frac{\dim\, ad\hspace{0.1pt}j}{\dim\, r}I_2^{(r)}.
\end{equation}

The quantities of interest for the present study are listed in Table~\ref{tab}; they can be easily computed from Tables V, XII, and XIII of~\cite{fuchs}. Some low-dimensional algebras have not been presented in Table~\ref{tab} but their properties follow from the isomorphisms A$_1\cong$ B$_1\cong$ C$_1\cong$ D$_1$, B$_2\cong$ C$_2$, D$_2\cong$ A$_1\oplus$ A$_1$, and D$_3\cong$ A$_3$. Notice the particular case of E$_8$ for which the adjoint dimension is also the lowest-dimensional one. For completeness, we mention that the compact real forms of A$_r$, B$_r$, C$_r$, and D$_r$ are respectively $\mathfrak{su}(r+1)$, $\mathfrak{so}(2r+1)$, $\mathfrak{sp}(r)$, and $\mathfrak{so}(2r)$.

\begin{table*}[t]
\setlength{\extrarowheight}{3pt}
\begin{center}
\begin{tabular}{l|l|l|c|c} 
\hline
             & $f$ ($\bar f$)      & $ad\hspace{0.1pt}j$       & $C_2^{(ad\hspace{0.1pt}j)}$ & $C_2^{(ad\hspace{0.1pt}j)}/C_2^{(f)}$ \\ \hline 
A$_1$        & ${\bm 2}=(1)$ & ${\bm 3}=(2)$ & 2           & 8/3 \\ 
A$_{r\geq2}$ & $\bm{ r+1}=(1,0\dots,0),$ & $\bm{ r^2+2r}=(1,0\dots,0,1)$ & $r+1$ & $\frac{2(r+1)^2}{r(r+2)}$ \\ 
             & $\bm{\overline{ r+1}}=(0,\dots,0,1)$ & &  &  \\   \hline 
B$_{r\geq3}$ & $\bm{2 r+1}=(1,0,\dots,0)$ & $\bm{2 r^2+r}=(0,1,0,\dots,0)$ & $2r-1$ & $\frac{2r-1}{r}$ \\ \hline 
C$_{r\geq 2}$ & $\bm{2 r}=(1,0,\dots,0)$ & $\bm{2 r^2+r}=(2,0,\dots,0)$ & $r+1$ & $\frac{4(r+1)}{2r+1}$\\ \hline 
D$_4$ & $\bm 8=(1,0,0,0)$ or & $\bm{28}=(0,1,0,0)$ & 6 & 12/7 \\
      & $\ \ \ (0,0,1,0)$ or $(0,0,0,1)$ &      & & \\
D$_{r\geq 5}$ & $\bm{2r}=(1,0,\dots,0)$ & $\bm{2r^2-r}=(0,1,0,\dots,0)$ & $2(r-1)$ & $\frac{4(r-1)}{2r-1}$ \\ \hline 
E$_6$ & $\bm{27}=(1,0,\dots,0),$ & $\bm{78}=(0,\dots,0,1)$ & 12 & 18/13 \\          
      & $ \overline{\bm{27}}=(0,\dots,1,0)$ &  &  &  \\ 
E$_7$ & $\bm{56}=(0,\dots,1,0)$ & $\bm{133}=(1,0,\dots,0)$ & 18 & 24/19 \\ 
E$_8$ & & $\bm{248}=(1,0,\dots,0)$ & 30 &  \\  \hline 
F$_4$ & $\bm{26}=(0,0,0,1)$ & $\bm{52}=(1,0,0,0)$ & 9 & 3/2 \\ \hline 
G$_2$ & $\bm{7}=(0,1)$ & $\bm{14}=(1,0)$ & 4 & 2 \\
\hline              
\end{tabular}
\end{center}

\caption{Data related to simple Lie algebras: Lowest-dimensional or fundamental representation, $f$ (second column), adjoint representation, $ad\hspace{0.1pt}j$ (third column), and eigenvalues of the quadratic Casimir operator in those representations (fourth and fifth columns). The dimension of each representation is written in bold and its Dynkin labels are given between parenthesis. When existing (for A$_{r\geq2}$ and E$_6$), the representation conjugate to $f$, $\bar f$, is also given. Conventions of \cite{fuchs} are followed.}
\label{tab}
\end{table*}

We finally refer the interested reader to \cite{lie}, in which a powerful computer algebra package for Lie group computations (especially tensor products) can be found. 

\section{Comments on ${\cal N}=1$ SUSY Yang-Mills}\label{susya}

The sign of the coefficients $\beta_0$ and $\beta_1$ are unaltered when one flavour of adjoint fermions is added~\cite{caswell} and the equalities (\ref{g2}) still hold, so one still expects confinement and bound states to appear in SUSY YM theory. It has also recently been shown that fermions in the adjoint representation of $\mathfrak{su}(3)$ gain a dynamically generated mass due to chiral symmetry breaking~\cite{aguilar0}. As for the quasigluons, this is an indication that a ``quasigluino" picture might be relevant as well in view of clarifying the structure of the low-lying SUSY YM spectrum. 

The quasigluino is a Majorana spinor with an intrinsic parity $\eta_{\tilde g}^2=-1$. A two-quasigluino bound state, also called gluinoball and denoted $\tilde g\tilde g$, must have a positive charge conjugation because the quasigluinos are self-conjugate. Moreover, the singlet made by the tensor product of two adjoint representations is symmetric for any gauge algebra, see (\ref{gluel}). The Pauli principle then imposes that $L+S$ is even in a standard $\left|^{2S+1}L_J\right\rangle$ basis \cite{ggsusy}, and one also obtains that $P=(-)^{L+1}$ for a $\tilde g \tilde g$ state. Thus the lightest gluinoball quantum state reads $\left|\tilde g\tilde g;0^{-+}\right\rangle=\left|^1S_0\right\rangle$.  Quasigluino-quasigluon bound states can also be formed, and the state presenting the dominant $S$-wave component, that is the lightest one, has spin $1/2$ and can be expressed as $\left[2/3\right]^{1/2}\left|^2 S_{1/2}\right\rangle-\left[1/3\right]^{1/2}\left|^4 D_{1/2}\right\rangle$. Since it has a nonzero $D$-wave component, it will be heavier than the pseudoscalar gluinoball, that we thus predict to be the lightest bound state of SUSY YM theory (gluelumps excepted). Notice that such a mass hierarchy relies on the assumption that the quasigluon and quasigluinos dynamically generated masses are similar, typically $O(\sqrt{\sigma_0})$. This is not in contradiction with the first estimates of \cite{aguilar0} that, however, suggest a larger value for the gluino mass, leading eventually to a reduction of the mass gap between the lightest gluinoball and glueball. An quantitative calculation of this mass gap would require to deal with an explicit model, and is out of the scope of the present work.

Note that the state $\left|\tilde g\tilde g;0^{-+}\right\rangle$ has been found to be the lightest one of the SUSY YM spectrum in \cite{feo} for the gauge algebra $\mathfrak{su}(N)$ by using large $N$ arguments. The pseudoscalar gluinoball has been found to be lighter than the scalar glueball also in the lattice study \cite{susylat}, where the SUSY YM theory with gauge algebra $\mathfrak{su}(2)$ is used.  

\end{appendix}

\end{document}